\documentclass{raa}          

\usepackage{graphicx,times,epsf}             
\usepackage{amsmath}
\usepackage{multirow}

\begin{document}

   \title{Gravitationally lensed extended sources:\\the case of QSO RXJ0911
}

   \volnopage{Vol.0 (200x) No.0, 000--000}      
   \setcounter{page}{1}          
   
      \author{D.T. Hoai\inst{1}
          \and
          P.T. Nhung\inst{1}
          \and
          P.T. Anh\inst{1,2}
          \and
          F. Boone\inst{2}
          \and
          P. Darriulat\inst{1}
          \and
          P.N. Diep\inst{1}
          \and
          P.N. Dong\inst{1}
          \and\\
          N.V. Hiep\inst{1}
          \and
          N.T. Thao\inst{1}
          }

   \institute{VATLY, INST, 179, Hoang Quoc Viet, Cau Giay, Hanoi, Vietnam \and
             IRAP, 14, avenue Edouard Belin, F31400 Toulouse, France
             }

   \date{}

\abstract{After a brief reminder of the mechanism of gravitational lensing of extended sources, the particular case of the host galaxy of QSO RXJ0911, a high redshift ($z \sim 2.8$) quadruply imaged quasar, is explored. The non linearity of the problem, together with the proximity of the source to a cusp of the lens inner caustic, have important consequences on the dependence of the image appearance on the size and shape of the source. Their expected main features and their interpretation in terms of source extension and shape are investigated in a spirit of simplicity and in preparation for the analysis of high sensitivity and spatial resolution images that will soon be within reach with the completion of the Atacama  Large Millimeter/submillimeter Array (ALMA). In particular, the information on source size carried by relative image brightness is discussed. Extension of the results to other types of quadruply imaged quasars is briefly considered.
}

   \authorrunning{D. T. Hoai et al.}            
   \titlerunning{}  

   \maketitle

%

\section{Introduction}

Strong gravitational lensing has become a textbook topic. In particular, several authors, such as Blandford et al.~(\cite{blan2}) or Saha and Williams~(\cite{saha}), have summarized the main properties in simple terms, underlining the most general qualitative features. While the case of complex lens configurations has been extensively studied (see for example Jullo et al.~(\cite{jull})), in particular with the aim of evaluating the mass distribution of baryonic and dark matter in cluster lenses, studies of strongly lensed extended sources are less common. The clearest cases of strong lensing, which are naturally the most studied, are often associated with sources located near the inner caustic of the lens, making the problem highly non linear: when crossing the caustic outward, magnifications become infinite and one switches from a four-image to a two-image configuration. Several authors, such as Dominik~(\cite{domi}), Bartelmann~(\cite{bart}), Suyu and Blandford~(\cite{suyu}) or Suyu et al.~(\cite{suyu2}) have analysed the consequences in the case of extended sources and described their effects in some detail. With the completion of high sensitivity and angular resolution instruments, such as the Atacama Large Millimeter/submillimeter Array (ALMA) or the Square Kilometer Array (SKA), the study of host galaxies of high redshift gravitationally lensed quasars will enter a new phase with the ability of resolving spatially the sources in both their molecular gas and dust contents.

The present work concentrates on the case of a quadruply imaged quasar, RXJ0911, which provides a good illustration of the main features. Observations in the visible and near-infrared by Burud et al.~(\cite{buru}), using the 2.56 m Nordic Optical Telescope and the ESO 3.5 m New Technology Telescope, resolved clearly the object in four QSO images having a redshift $z \sim 2.8$ and an elongated lens galaxy and measured the relative image fluxes, their magnitudes and positions. Three of the four images, named A1 to A3, are at similar distances from the lens ($\sim 0.9\arcsec$) while the fourth image, B, is 2.5 times as far ($\sim 2.2\arcsec$). Modelling such lensing properties (Burud et al.~\cite{buru}) requires the presence of an external shear in addition to the lens ellipticity. Two years later, the lens was identified by Kneib et al.~(\cite{knei}) as being a member of a galaxy cluster having a mean redshift $z = 0.769 \pm 0.002$ and a velocity dispersion of 836 $\pm$ 190 km s$^{-1}$ using the Low Resolution Imaging Spectrograph of the Keck II Telescope. The cluster has an estimated mass of $\sim (6.2 \pm 2.7) 10^{14}$ solar masses. 

Recently, RXJ0911 has been observed using the Plateau de Bure Interferometer (Tuan Anh et al.~\cite{tuan}) tuned in the CO(7-6) region. At such wavelengths, the source is no longer the quasar, which can be considered as a point source, but its host galaxy. More precisely, the CO line is associated with the presence of molecular gas and the continuum with that of dust: measuring the source extension in each case would provide important information on the physics of early galaxies. The Plateau de Bure data resolve the source spatially on the CO line, suggesting a root mean square radius of several hundred parsecs ($1\arcsec$ in the sky corresponds to 8 kpc in the source plane). The evidence for resolving spatially the source in the continuum is much weaker, at the level of only two standard deviations. Such results exploit the performance of the radio telescope array to the limits of its ability but give confidence that the better acceptance, resolution and sensitivity of ALMA, soon to be completed, will give more definitive results. The aim of the present work is to describe in simple terms what can be expected in order to guide the analysis of future higher resolution data.

\begin{figure}
   \centering
   \includegraphics[width=0.35\textwidth]{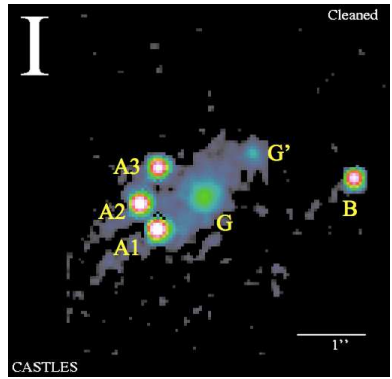}
   \caption{Image configuration (Kochanek~\cite{koch}) of RXJ0911 showing the lensing galaxies G and G'.}
              \label{Fig1}
   \end{figure}

A HST image of the quasar is shown in Figure~\ref{Fig1}, the associated data are listed in Table 1.

%

\begin{table*}
\caption{Position data from CASTLES (upper line, Kochanek et al.~\cite{koch}) and best fit results from Burud et al.~(\cite{buru}), lower line. Magnifications (M) are from Burud et al.~(\cite{buru}) model.}             
\label{table1}      
\centering        
\begin{tabular}{|c|c|c|c|c|c|c|c|}
\hline
\multicolumn{2}{|c|}{Observations} & A1 & A2 & A3 & B & G & G'\\
\cline{1-8}
\multicolumn{1}{|c}{\multirow{4}{*}{Position}} &
\multicolumn{1}{|c}{\multirow{2}{*}{--RA(arcsec)}} &
\multicolumn{1}{|c|}{0} & $-0.260$ & 0.018 & 2.972 & $0.698$ & $1.452$ \\ 
\multicolumn{1}{|c}{}                        &
\multicolumn{1}{|c}{}                        &
\multicolumn{1}{|c|}{0} & $-0.259$ & 0.013 & 2.935 & - & -     \\ \cline{2-8}
\multicolumn{1}{|c}{}                        &
\multicolumn{1}{|c}{\multirow{2}{*}{Dec(arcsec)}} &
\multicolumn{1}{|c|}{0} & 0.406 & 0.960 & 0.792 & 0.512 & 1.177   \\ 
\multicolumn{1}{|c}{}                        &
\multicolumn{1}{|c}{}                        &
\multicolumn{1}{|c|}{0} & 0.402 & 0.946 & 0.785 & - & -    \\ \cline{1-8}
M & & $-4.45$ & 8.59 & $-3.70$ & 1.79 & - & - \\
\hline
 \end{tabular}
\end{table*}

\section{Strong lensing: a reminder}
\subsection{General formalism}
Fermat principle states that images form where the gradient of the time delay, $\tau$, cancels. In the approximation of small deflections, which always applies in practice, the time delay can be written as the sum of a geometrical delay and of the gravitational delay proper: $\tau = \tau_{0}[{1/2} (\textbf{\emph{i}} - \textbf{\emph{s}})^2 - \psi]$. Here, \textbf{\emph{i}} and \textbf{\emph{s}} are the image and source vectors in sky coordinates (in a plane normal to the line of sight), $\tau_0$ is a constant time scale and $\psi$ is an effective potential that describes the deflection induced by the lens as a function of the sky coordinates of the image. The effective potential $\psi$ is proportional to the integral of the gravity potential along the line of sight between source and observer. A convenient form, used by many authors, includes an elliptical lens and an external shear, the axes of the ellipse being taken as coordinate axes without loss of generality: 
\begin{equation}
      \psi = r_0 r (1+\varepsilon \mathrm{cos}2\varphi)^{1/2} + 1/2 \gamma_0 r ^2 \mathrm{cos}2(\varphi - \varphi_0)  \,,
   \end{equation}	
where $(r,\varphi)$ are the polar coordinates of the image. The potential is usually given in Cartesian coordinates, as in Burud et al.~(\cite{buru}); we rewrote it here in polar coordinates, which we found convenient in the case where the potential takes such a simple form, but this is a technical detail of no relevance to the results of the study. The lens term, of strength $r_0$ and aspect ratio $[(1+\varepsilon)/(1 - \varepsilon)]^{1/2}$, decreases as $1/r$ outside the core region. The shear term has a strength $\gamma_0$ and makes an angle $\varphi_0$ with the major axis of the lens ellipse.  Writing that the gradient of Relation 1 cancels, and calling $(r_s,\varphi_s)$ the polar coordinates of the source, one obtains the lens equation:
\begin{equation}
      r_s e^{i \varphi s} = r e^{i \varphi} (1 - r ^{- 1} \partial{\psi} / \partial{r} - i r ^{- 2} \partial{\psi} / \partial{\varphi}) \,,
   \end{equation}	
As is well known, there may typically be two or four images depending on the position of the source with respect to the inner caustic of the lens. If the potential is isotropic, the lens equation reduces to $ r_s e^{i \varphi s}=e^{i \varphi} (r - \partial{\psi} / \partial{r})$, which has two obvious solutions, one at $\varphi_+=\varphi_s$ and the other at $\varphi_-=\varphi_s +\pi$ with $r_+=\partial{\psi}/\partial{r}+r_s$ and $r_{-}=\partial{\psi}/\partial{r} - r_s$ respectively. For $r_s=0$, the alignment is perfect and one obtains an Einstein ring having $r=\partial{\psi} / \partial{r}$. 

\subsection{Extended sources} 
To the extent that the source is small and not too close to the lens inner caustic, the image of an extended source is simply obtained by differentiating the lens equation, 
$r_s e^{i \varphi s}=Dre^{i \varphi}$, where $D=D_r+iD_i$ and $D_r=1-r^{- 1}\partial{\psi}/\partial{r}$, $D_i=-r^{- 2}\partial{\psi}/\partial{\varphi}$ . We obtain this way the relation between a point \mbox{$(r_s+dr_s,\varphi_s+d\varphi_s)$} on the source and its image $(r+dr,\varphi+d\varphi)$:
\begin{equation}
(dr_s+ir_s d\varphi_s)e^{i\varphi s}=D(dr+ird\varphi)e^{i\varphi}+
(\frac{\partial{D}}{\partial{r}}dr+\frac{\partial{D}}{\partial{\varphi}}d\varphi)re^{i\varphi} 
\end{equation}
In practice, one does not directly observe the source, but only its images. It is therefore convenient to replace, in the left-hand side of the above relation, the source dependent term $r_s e^{i\varphi s}$ by its expression in terms of the image coordinates: 

\begin{equation}
D(\frac{dr_s}{r_s}+id\varphi_s)=(D+r\frac{\partial{D}}{\partial{r}})\frac{dr}{r}+
(iD+\frac{\partial{D}}{\partial{\varphi}})d\varphi
\end{equation}

 \begin{figure*}
   \centering
   \includegraphics[width=0.8\textwidth]{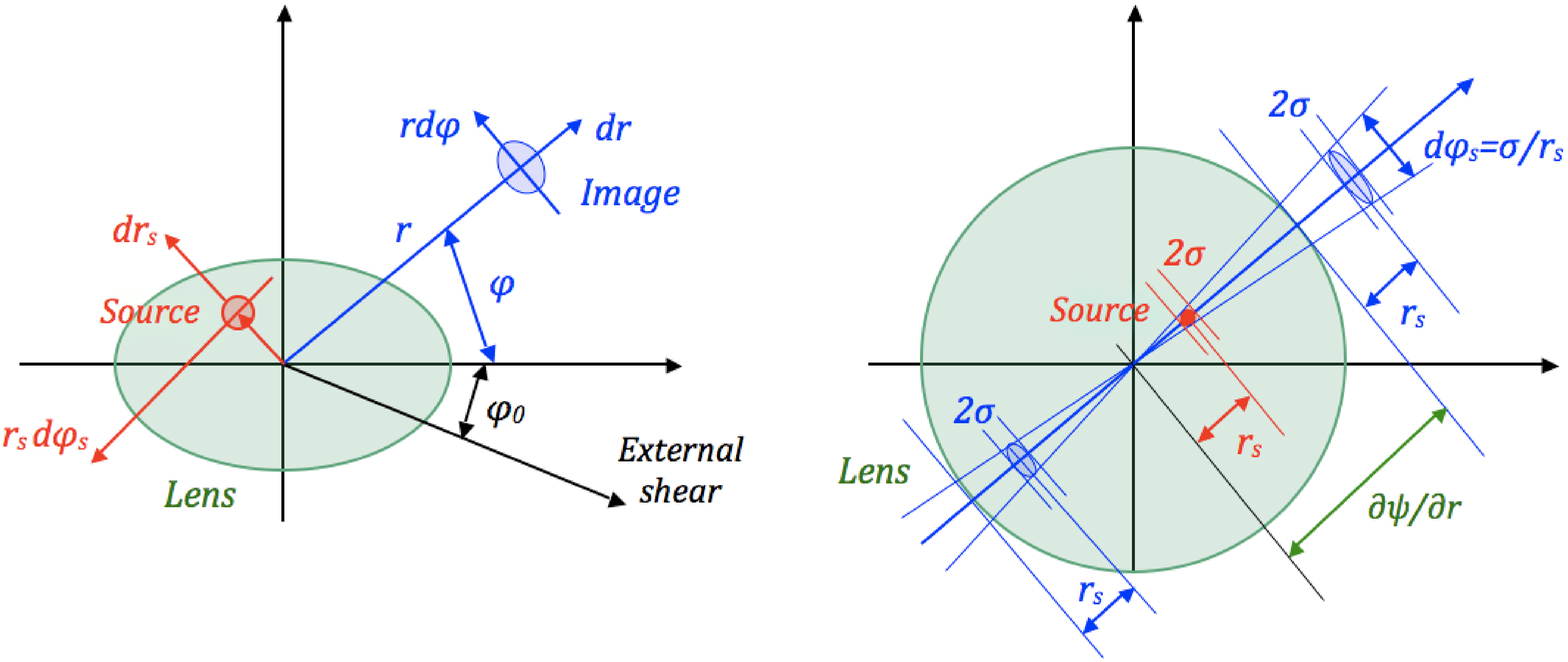}
   \caption{Left panel: Schematic geometry and definition of coordinates. The origin is the centre of lens G, the polar axis is positive westward and angles are measured counter clockwise. Right panel: Isotropic potential imaging.}
              \label{Fig2}
   \end{figure*}

Relation 3 gives the coordinates of an image point as a function of those of the corresponding source point. Indeed, $dr_s$ and $r_s d\varphi_s$ are Cartesian coordinates having their origin at the centre of the source and the axis of abscissas radial outwards; similarly, $dr$ and $rd \varphi$ are Cartesian coordinates having their origin at the centre of the image and the axis of abscissas radial outwards (Figure~\ref{Fig2}). For Relation 3 to apply, $dr_s$ and $r_sd \varphi_s$ must be small enough for the corresponding source points to stay away from the lens inner caustic. In practical cases, with the centre of the source located near the caustic, this will often not be the case if the source extension is such that it overlaps the caustic. The linear approximation of Relation 3 needs therefore to be used with care. Yet, it usefully serves several purposes, such as providing explicit expressions for the magnifications  which may be calculated using arbitrarily small values of $dr_s$ and $r_sd \varphi_s$ -- or for giving a qualitative illustration of the main features in simple terms as is done below. Relation 3 being linear, it is straightforward to express $(dr/r,d \varphi)$ as a function of $(dr_s/r_s,d \varphi_s)$ in terms of a lensing matrix $\lambda$ defined as:

\begin{subequations}
\begin{equation}
dr/r=\lambda_{11}dr_s/r_s+\lambda_{12}d\varphi_s\\
\end{equation}
\begin{equation}
d\varphi=\lambda_{21}dr_s/r_s+\lambda_{22}d\varphi_s
\end{equation}
\end{subequations}

with:
\begin{equation}
\begin{array}{l}
\lambda_{11}=[D_r(D_r+{\partial{D_i}}/{\partial{\varphi}})+
	D_i(D_i-{\partial{D_r}}/{\partial{\varphi}})]/ {\mu} \\
\lambda_{12}=- (D_r {\partial{D_r}}/{\partial{\varphi}}+
	D_i {\partial{D_i}}/{\partial{\varphi}})/ {\mu} \\				
\lambda_{21}=r (D_i {\partial{D_r}}/{\partial{r}}-
	D_r {\partial{D_i}}/{\partial{r}}])/ {\mu}	\\				        
\lambda_{22}=[D_r (D_r+r {\partial{D_r}}/{\partial{r}})+
	D_i (D_i + r {\partial{D_i}}/{\partial{r}})] /{\mu} \\				
\mu=(D_r+ r {\partial{D_r}}/{\partial{r}})( D_r+ {\partial{D_i}}/{\partial{\varphi}})+\\
\quad \quad (D_i+ r {\partial{D_i}}/{\partial{r}})( D_i - {\partial{D_r}}/{\partial{\varphi}})
\end{array}
\end{equation}

For an isotropic source brightness, such that $<dr_s/r_s>=0$, $<d\varphi_s>=0$, $< d\varphi_s  dr_s/r_s>=0$ and $<dr{_s^2}>=r{_s^2}<d\varphi{_s^2}>=\sigma^2$, it is straightforward to write the expressions of the semi-major and semi-minor axes $a$ and $b$ of the image, its position angle $\theta$ and magnification $M$:
\begin{equation}
\begin{array}{l}
(a\pm b)=(\sigma r/r_s)[|\lambda|^2 \pm 2 det(\lambda)]^{1/2}\\
\mathrm{tan}2\theta=(\lambda_{11}\lambda_{21}+\lambda_{12}\lambda_{22})/
(\lambda{_{11} ^2}+\lambda{_{12} ^2}-\lambda{_{21} ^2}-\lambda{_{22} ^2})\\
M=(r/r_s)^2 det(\lambda)\,,
\end{array}
\end{equation}
where $det(\lambda)=\lambda_{11}\lambda_{22}-\lambda_{12}\lambda_{21}$ and $|\lambda|^2=\lambda{_{11}^2}+\lambda{_{12} ^2}+\lambda{_{21} ^2}+\lambda{_{22} ^2}$.\\

The case of an isotropic potential (Figure~\ref{Fig2}, right) illustrates the above relations, with $tan2\theta=0$, $a = \sigma(1-d^2\psi/dr^2)^{-1}$ and $b = \sigma r/r_s$. The image is stretched normally to the lens-image direction by a factor $r/r_s$. The magnification in the lens-image direction is unity when the second derivative of the potential cancels. When small perturbations are added to the isotropic potential in the form of an external shear and/or a non-zero eccentricity of the lens, $a$, $b$ and $\theta$ deviate by correspondingly small amounts from the above values. When $r_s$ decreases, the two images get more and more elongated in the tangential direction forming characteristic arcs and finally merge while drifting toward the Einstein ring of radius $\partial \psi/ \partial r$.

The case of a potential $\psi = r r_0 (1+\varepsilon \mathrm{cos}2\varphi)^{1/2}$
that describes an elliptical lens without external shear, $\varepsilon$ accounting for the ellipticity of the lens, is particularly simple: one then has $<dr^2>=\sigma^2$, \mbox{$<r^2 d\varphi^2>=\sigma^2 / \mu^2$} and $<r dr d\varphi>=0$. The images are ellipses stretched tangentially, their minor axes are equal to the source diameter and, writing $ \mu=1-r_0 r^{-1} (1-\varepsilon^2)(1+\varepsilon \mathrm{cos}2\varphi)^{-3/2} $, their major axes are a factor $1/\mu$ larger. At constant $\varphi$, $1/\mu$ is of the form $r/(r-k r_0)$ with $ k = (1-\varepsilon^2)(1+\varepsilon \mathrm{cos}2\varphi)^{-3/2} $: it reaches unity at large values of $r$ and increases when $r$ decreases toward the critical curve, of equation $r=kr_0$, where $1/\mu$ diverges. 

\begin{figure*}
   \centering
   \includegraphics[width=0.8\textwidth]{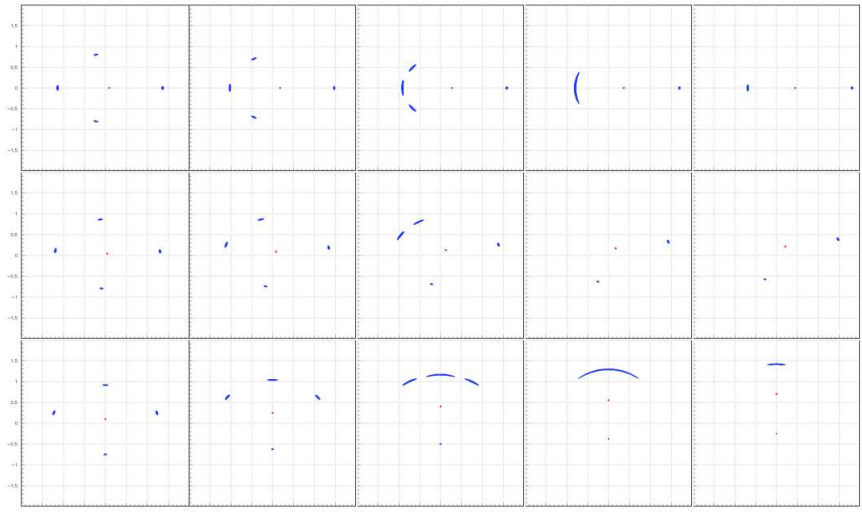}
   \caption{Dependence of the image positions and morphology on the source position. The lens is circular ($r_0=1\arcsec$) and centred at the origin. The potential includes an external shear, $\gamma_0=0.2$, $\varphi_0=0$. From left to right: $\varphi_s=0\degr$, $r_s= 0.09\arcsec, 0.18\arcsec, 0.27\arcsec, 0.36\arcsec, 0.45\arcsec$ (upper panels); $\varphi_s=45\degr$, $r_s= 0.06\arcsec, 0.12\arcsec, 0.18\arcsec, 0.24\arcsec, 0.30\arcsec$ (second row); $\varphi_s=90\degr$, $r_s= 0.10\arcsec, 0.25\arcsec, 0.40\arcsec, 0.55\arcsec, 0.70\arcsec$ (lower panels). The source is a disk of radius $0.02\arcsec$. }
              \label{Fig3}
   \end{figure*}

\begin{figure*}
   \centering
   \includegraphics[width=0.75\textwidth]{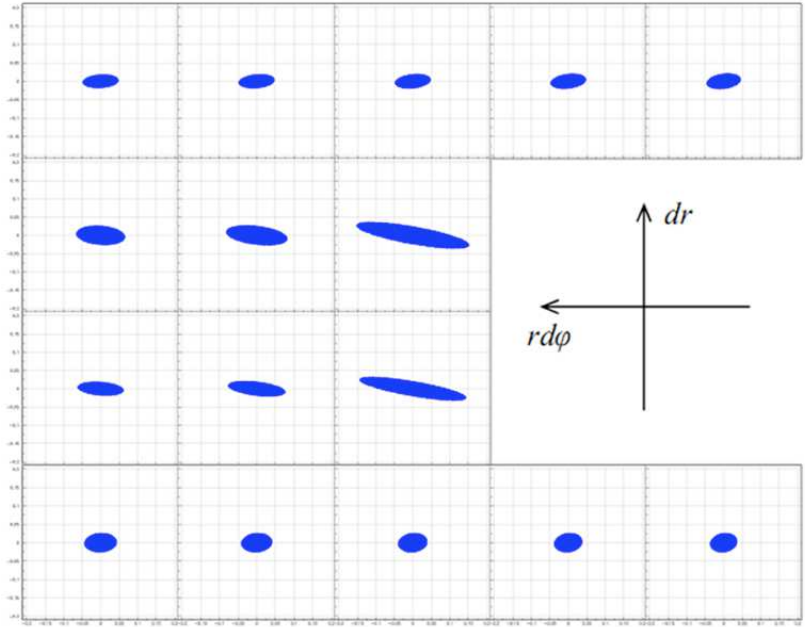}
   \caption{Image appearances for the same potential and source size as in Figure~\ref{Fig3}. Source positions are: $\varphi_s=20\degr$, $r_s= 0.1\arcsec, 0.15\arcsec, 0.2\arcsec, 0.25\arcsec, 0.3\arcsec$ from left to right. Images are in [$dr$ (up), $r d\varphi$ (left)] coordinates. Images 1 to 4, numbered from South clockwise, are displayed in the upper to lower rows respectively.}
              \label{Fig4}
   \end{figure*}

Similar results are obtained for a potential of the form $\psi = r_0 r + 1/2 \gamma_0 r^2 \mathrm{cos}2\varphi $ that describes a circular lens with external shear along the axis of abscissas. Figure~\ref{Fig3} shows how the positions of the images vary as a function of the source position: one switches from a four-image configuration, when the source is inside the caustic, to a two-image configuration, when the source is outside the caustic. Image shapes, illustrated in Figure~\ref{Fig4} for different source positions, are again ellipses stretched tangentially but, at variance with the preceding case, they are slightly tilted as soon as $\varphi_s$ deviates from $0\degr$ or $90\degr$. Qualitatively, for small values of the external shear, the general features of the isotropic case are maintained.

%

\section{QSO RX J0911} 
\subsection{Quasar point images}
We study the image configuration of RXJ0911 using one of two potentials: 
\begin{equation}
    \psi_1 = r_0 r [1+\varepsilon \mathrm{cos}2(\varphi- \varphi_1)]^{1/2}
    + 1/2 \gamma_0 r^2 \mathrm{cos}2(\varphi- \varphi_0)
\end{equation}
used by many authors (Blandford \& Kochanek~\cite{blan}, Saha \& Williams~\cite{saha}, Peeples~\cite{peep}, Burud et al.~(\cite{buru}), Kassiola \& Kovner~\cite{kass}, Witt \& Mao~\cite{witt}, Kneib et al.~\cite{knei}) and including an elliptical main lens G and an external shear, or
\begin{equation}
    \psi_2 = r_0 r + r'_0 [{r^{*}}^2+r^2- 2rr^{*} \mathrm{cos}(\varphi- \varphi^{*})]^{1/2}
    + 1/2 \gamma_0 r^2 \mathrm{cos}2(\varphi- \varphi_0) 
\end{equation}
describing a circular main lens G, a small perturbation from its satellite galaxy G' and an external shear [Kochanek~\cite{koch2}, Schechter~\cite{sche}]. 

\begin{figure}
   \centering
   \includegraphics[width=0.4\textwidth]{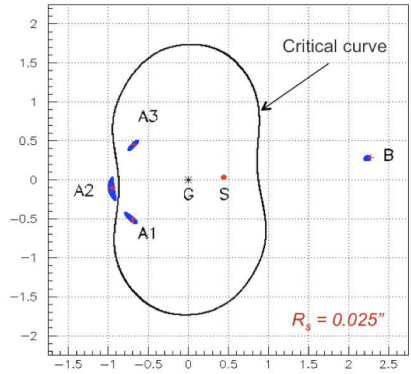}
   \caption{Image positions obtained using potential $\psi_1$ are shown in blue for a source radius of $0.025\arcsec$. The source is shown in red. The lens G is at the origin of coordinates (black asterisk). Red crosses indicate the HST observed image positions. The results obtained using potential $\psi_2$ are identical within the precision of the figure. Sky coordinates are in arcsec.}
              \label{Fig5}
   \end{figure}

The parameters of the model are adjusted by minimizing the $\chi^2$ describing the match between the observed quadruple point images and the prediction of the point source model. The reason for using two different potentials is to illustrate the relative independence of most of the main results on the detailed form of the potential, not to argue about which form is better. Care has been taken, when exploring the $\chi^2$ dependence on the coordinates of the point source, to search for minima confined within the lens inner caustic. This precaution is important when the minimization algorithm approximates locally the surface of constant $\chi^2$ by a quadratic form in order to evaluate the gradient along which to move and to calculate uncertainties. The results are summarized in Tables 2 and 3. Images obtained using potential $\psi_1$ for a disk source of uniform brightness over a radius $R_s=0.025\arcsec$ are shown in Figure~\ref{Fig5}, together with the critical curve. The images obtained using potential $\psi_2$ are practically indistinguishable from those obtained using potential $\psi_1$.

\begin{table}
\caption{Best fit results ($r$ in arcseconds and $\varphi$ in degrees).}             
\label{table2}
\centering        
\begin{tabular}{|c|c|c|c|c|}
\hline
	& A1 & A2 & A3 & B\\
\hline
$r_{obs}$ & 0.866 & 0.964 & 0.814 & 2.291\\
$r_{model} (\psi_1)$ & 0.853 & 0.968 & 0.819 & 2.294\\
$r_{model} (\psi_2)$ & 0.845 & 0.967 & 0.828 & 2.282\\ 
\hline
$\varphi_{obs}$ & 216.3 & 186.3 & 146.6 & 7.0\\
$\varphi_{model} (\psi_1)$ & 219.3 & 188.2 & 150.0 & 5.8\\
$\varphi_{model} (\psi_2)$ & 219.0 & 188.4 & 151.2 & 6.2\\
\hline
M ($\psi_1$) & $-4.14$ & 8.33 & $-3.37$ & 1.92\\
M ($\psi_2$) & $-4.45$ & 8.32 & $-3.47$ & 1.76\\
\hline
 \end{tabular}
\end{table}
%
  
\begin{table}
\caption{Best fit values of the parameters (angles in degrees and angular distances in arcseconds).}             
\label{table3}
\centering        
\begin{tabular}{|c|c|c|c|c|c|c|}
\hline
\multicolumn{7}{|c|}{Potential $\psi_1$} \\
\hline
$r_0$ & $\varepsilon$ & $\varphi_1$ & $\gamma_0$ & $\varphi_0$ & $r_s$ & $\varphi_s$ \\
\hline
1.1085 & $-0.0237$& 65.0 & 0.309 & 7.32 & 0.4456 & 3.84 \\
\hline \hline
\multicolumn{7}{|c|}{Potential $\psi_2$ ($r*=1.005$, $\varphi*=41.4$)} \\
\hline 
$r_0$ & $r'_0$ & - & $\gamma_0$ & $\varphi_0$ & $r_s$ & $\varphi_s$ \\
\hline 
1.0931 & 0.01935 & - & 0.317 & 6.7 & 0.4500 & 6.7 \\
\hline
\end{tabular}
\end{table}
%

\subsection{Extended source}
\begin{figure*}
   \centering
   \includegraphics[width=0.6\textwidth]{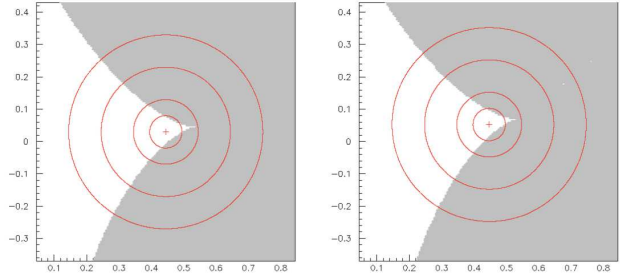}
   \caption{Position of the source with respect to the caustic for potentials $\psi_1$ (left panel) and $\psi_2$ (right panel). The concentric circles are for $R_s = 0.05\arcsec, 0.1\arcsec, 0.2\arcsec$ and $0.3\arcsec$. The white areas show the inner region of the caustic where the source needs to be located to form four images. When the source is outside this region only two images are formed. Sky coordinates are in arcsec.}
              \label{Fig6}
   \end{figure*}

\begin{figure*}
   \centering
   \includegraphics[width=0.95\textwidth]{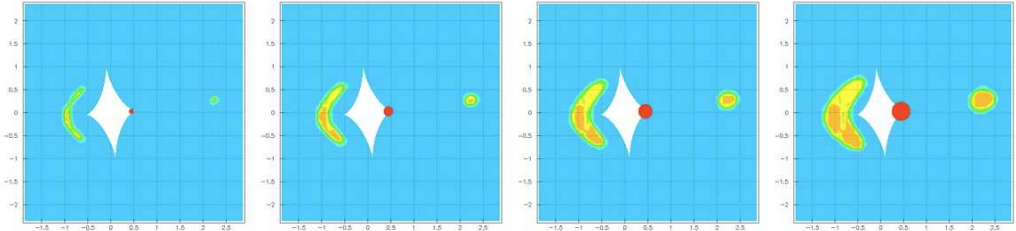}
   \caption{Images of RXJ0911 (assuming a disk source of radius $R_s = 0.05\arcsec, 0.1\arcsec, 0.15\arcsec$ and $0.2\arcsec$ from left to right). The white area shows the caustic and the source is shown as a red disk. Sky coordinates are in arcsec.}
              \label{Fig7}
   \end{figure*}

As can be seen from Table 1, the distance between the centres of image A2 and of its neighbours is $\sim0.6 \arcsec$ and the sums of the semi-major axes of A2 and A1 or A3 is $\sim9$ source radii: as soon as the source radius exceeds $\sim0.06 \arcsec$ to $0.07 \arcsec$, the three A images merge and the linear approximation developed in the preceding section is no longer appropriate. In the present section, images are therefore constructed point-by-point, each image point being given a weight equal to the absolute value of its magnification. In practice, we calculated, for each point of a fine-mesh map of the source plane, the positions and magnifications of its two or four image points in the image plane and we constructed images of a set of source points as the set of images of these source points, each time using the proper weight (including magnification and, in case of a non-uniform distribution, the source brightness). Both potentials give very similar results (Figure~\ref{Fig6}), with the source covering the western cusp of the caustic: the inner part of it gives four images, the outer part, only two. Indeed, as commented earlier by Saha \& Williams~(\cite{saha}), the precise position of the point images, including the strong asymmetry between the A and B images, severely constrains the position of the source with respect to the caustic cusp, with little dependence on the precise form of the effective potential. Uniform (radius $R_s$) and Gaussian ($\sigma_s$) brightness distributions are also found to give similar results in terms of $\rho_s=\sqrt{<r{_s^2}>}=R_s/\sqrt{2}=\sigma_s\sqrt{2}$. Figure~\ref{Fig7} displays the images obtained using potential $\psi_1$ for different source extensions.

%

The relative occurrence of two-image configurations increases with the size of the source, causing the A images to become globally fainter with respect to image B, the latter being always present in the two-image configuration. This suggests using the apparent integrated flux ratio, B/A, as a measure of the source size. This ratio is essentially unaffected by beam convolution and its measurement does not require a very good angular resolution (yet sufficient to separate A from B). However, as image B is not much magnified, it requires a high sensitivity to be obtained with good precision. 

\begin{figure*}
   \centering
   \includegraphics[width=0.85\textwidth]{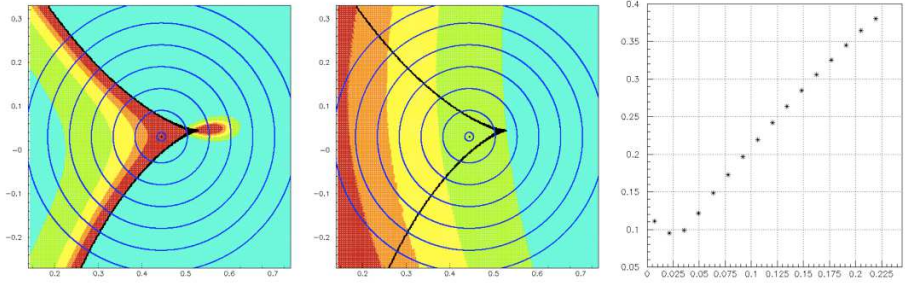}
   \caption{Distribution of the magnifications in the source plane for the three A images considered globally (left panel) and for image B (middle panel). Concentric circles show sources of respective radii $R_s= 0.01\arcsec, 0.06\arcsec, 0.11\arcsec, 0.16\arcsec, 0.21\arcsec, 0.26\arcsec$ and $0.31\arcsec$. In the left panel, from left to right, contours are for M = 6, 8, 10 and 12. In the middle panel, again from left to right, contours are for M = 2.0, 1.9, 1.8, 1.7 and 1.6. The right panel displays the dependence of the B/A integrated flux ratio (ordinate) on $\rho_s=R_s/\sqrt{2}$ (arcseconds) for a disk source of uniform brightness. One arcsecond corresponds to 8 kpc in the source plane.}
              \label{Fig8}
   \end{figure*}

To understand what happens, it is important to keep in mind that magnifications depend strongly on the position of the source point with respect to the caustic cusp, or equivalently on the position of the image point with respect to the critical curve. This is illustrated in Figure~\ref{Fig8}, which displays the variations over the source plane of the A and B magnifications separately.

The A images (considered here globally, namely the flux being integrated over the area covered by the three A images) are faint in most of the two-image region, large magnifications being only reached in the vicinity of the caustic cusp. On the contrary, the magnification of image B, which is far from the critical curve, is always weak, whether the source point is inside or outside the caustic cusp: it varies very slowly across the region explored in Figure~\ref{Fig8}.  When the source size increases, starting from a B/A integrated flux ratio of $\sim$11\% for a point source, as soon as the image size overlaps sufficiently the caustic, the B integrated flux grows essentially in proportion with the source area, the average magnification remaining between 1.5 and 2.2. But the relative growth of the A integrated flux is much slower: outside the caustic cusp, there is only one A image left and the magnification decreases rapidly when moving away from the cusp. Hence an increase of the B/A integrated flux ratio as a function of source size as displayed in the right panel of Figure~\ref{Fig8}.  

\subsection{Different configurations}
The above comments apply, \textit{mutatis mutandis}, to other cases of quadruply imaged quasars. Using the classification proposed by Saha \& Williams~(\cite{saha}) we briefly consider a few typical configurations: a \textquotedblleft{core quad}\textquotedblright, such as Q2237+030, H1413+117 (the Cloverleaf) or HST 1411+521, with the source close to the origin; an \textquotedblleft{inclined quad}\textquotedblright, such as MG0414+0534 or B1608+656, with the source close to the caustic but far from its cusps; a \textquotedblleft{long axis quad}\textquotedblright, such as B1422+231, with the source near a major axis cusp of the caustic. The case of RXJ0911, studied here in detail, is a \textquotedblleft{short axis quad}\textquotedblright{} with the source near the minor axis cusp. For each of these typical configurations, we use potential $\psi_1$ with the parameters taking the same values as for RXJ0911 (Table 2) and simply change the position of the source. 

\begin{figure*}
   \centering
   \includegraphics[width=0.85\textwidth]{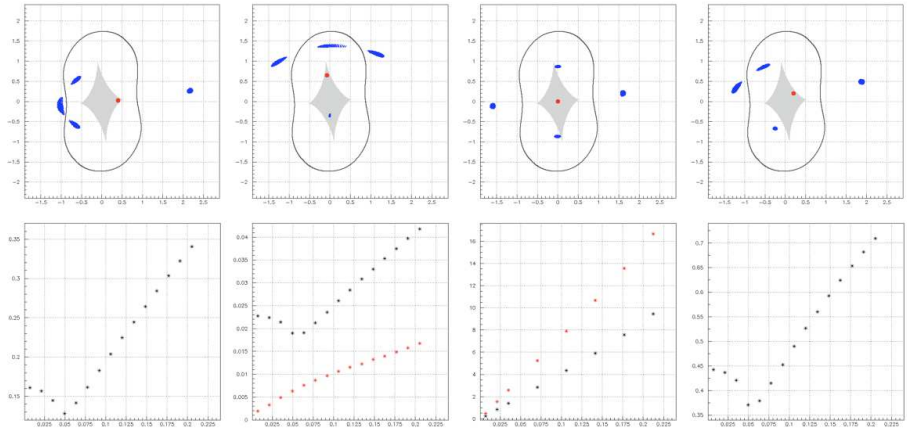}
   \caption{Typical quadruply imaged quasars according to Saha \& Williams~(\cite{saha}) classification. From left to right: short axis quad, long axis quad, core quad and inclined quad. The images (blue areas) are displayed in the upper panels for a disk source (red disk) having a radius $R_s= 0.05\arcsec$; the caustic is shown as a grey area. The lower panels illustrate the dependence of image integrated flux on source size. The two left panels display the dependence of the B/A integrated flux ratio on $\rho_s$ with distances from the source centre to the cusp of $0.11\arcsec$ for the short axis and $0.35\arcsec$ (black) and $0.07\arcsec$ (red) for the long axis. The core quad panel displays the mean radial size of the images (NS in red and EW in black) as a function of $\rho_s$. The right panel displays the brightness ratio of the south-western pair to the north-eastern pair.}
              \label{Fig9}
   \end{figure*}

The results are displayed in Figure~\ref{Fig9}. In the case of a long axis quad, as in the present case of a short axis quad, the B/A integrated flux ratio provides a good measure of the source size, however with a much lower magnification of image B, resulting from its being far from the critical curve: in such a long axis case, both A and B magnifications depend strongly on the distance of the source to the cusp. In the short axis case of Figure~\ref{Fig9}, this distance is $0.11\arcsec$ while in the long axis case, two values, $0.07\arcsec$ and $0.35\arcsec$ are displayed. In the case of a core quad, the source does not overlap the caustic, even at the maximal value of $\rho_s$ and the images do not merge: their relative brightness is almost independent from the source size but their radial extensions provide a good measure of it. Diametrically opposite images have nearly equal magnifications that are averaged in Figure~\ref{Fig9}. 

In the inclined quad case, the two north-eastern images merge as soon as the source overlaps the caustic and the ratio of the global brightness of the south-western images to that of the north-eastern images in displayed in Figure~\ref{Fig9}.
From this analysis we conclude that integrated flux ratios provide a useful measure of the source size in configurations where it is located in the vicinity of the inner caustic. However, when it is far from it, in configurations that are closer to an Einstein ring as is the case for the Cloverleaf, image extensions must be preferred to integrated flux ratios to evaluate the source size.

\subsection{Velocity gradients}

 \begin{figure*}
   \centering
   \includegraphics[width=0.85\textwidth]{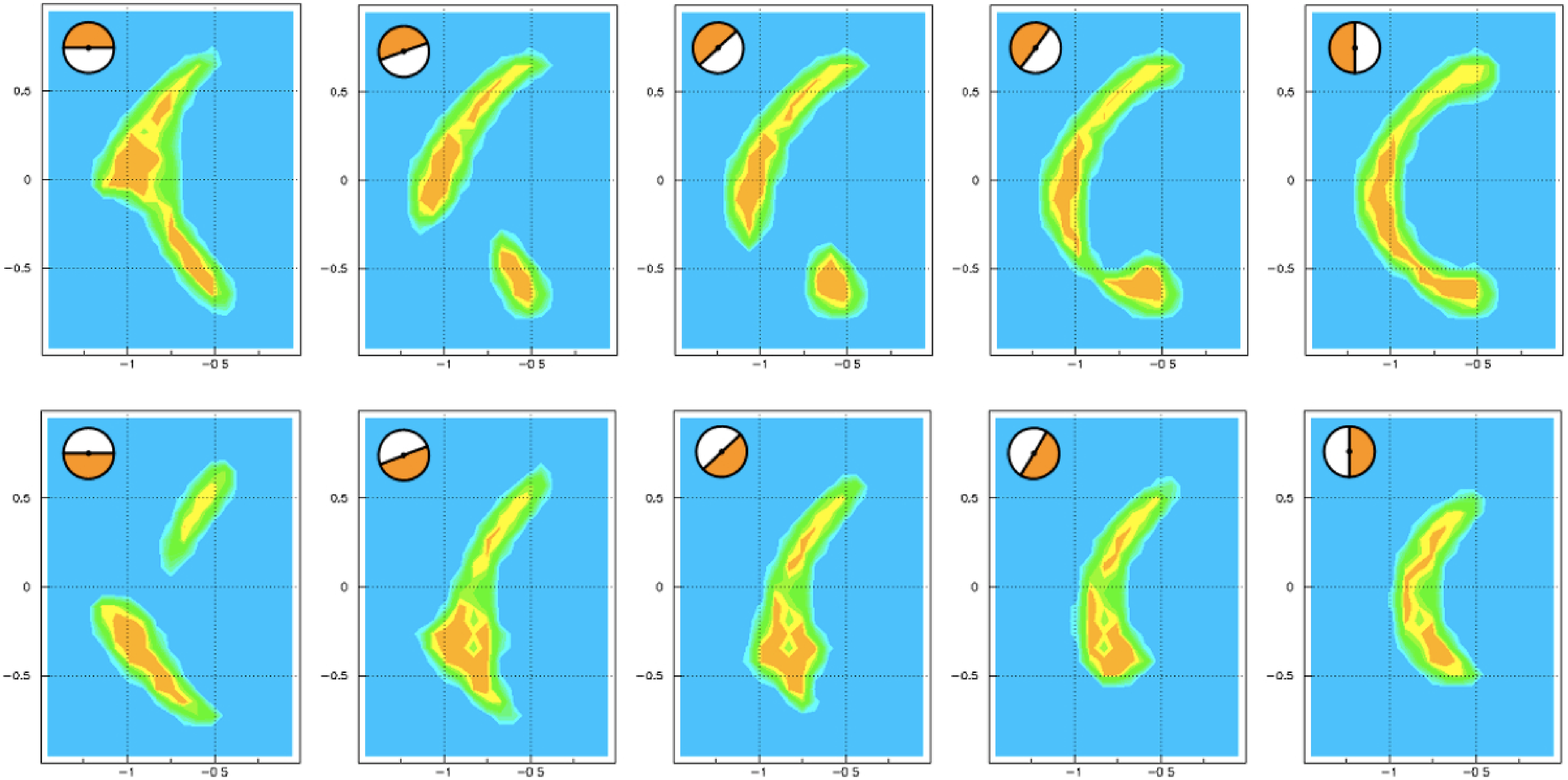}
   \caption{A images of a split source having $R_s=0.15\arcsec$. The split is along a diameter with position angles of 0\degr, 30\degr, 45\degr, 60\degr and 90\degr (from left to right). The upper panels are for the eastern half and the lower panels for the western half of the source.}
              \label{Fig10}
   \end{figure*}

Finally, we briefly comment on the effect of a velocity gradient on the appearance of the A images. Such gradient can be expected, for example, from a movement of rotation of the source or from molecular outflow. In the absence of gravitational lensing, the maps of the parts of the source associated with the red-shifted and blue-shifted parts of the velocity spectrum are obtained directly. However, in the presence of gravitational lensing, such is not the case when the source is extended. Reconstructing the maps of an extended source in general, and in particular of the parts of the source associated with the red-shifted and blue-shifted parts of the velocity spectrum, is more involved: one starts with some hypotheses about these maps, reconstructs the associated images and improves the hypotheses until one obtains a good match with the observed images. This iteration process is usually implemented by assuming some model for the source maps, the parameters of which are adjusted by minimizing the $\chi^2$ that measures the quality of the match between the observed and predicted images. At the end, if the models used for the blue-shifted and red-shifted parts of the source are adequate, the result is the same as in the absence of gravitational lensing. 

We illustrate this point in Figure~\ref{Fig10}, in the case of a disk source having $R_s=0.15\arcsec$ and split in two halves by a diameter making different angles with the east-west direction. In the absence of gravitational lensing, each half disk of the source would have as image a similar half disk while in the presence of gravitational lensing, the images are distorted and do not provide a direct picture of the source. In particular, in the present case of three merging images, A2, which has a magnification about twice those of A1 and A3, is inverted with respect to A1 and A3, increasing the importance of distortion. This complication, inherent to the gravitational lensing of extended sources, is a minor difficulty in comparison with the important benefit offered by the large magnifications associated with gravitational lensing, without which the source could not even be resolved in most cases.

\section{Summary}
The mapping of an extended source onto its gravitationally lensed images has been studied with the aim of guiding the analysis of future high resolution observations of the host galaxy of QSO RXJ0911 and of similar high redshift quasars: the results are general enough to be of interest to the study of other multiply imaged galaxies. Using simple effective potentials, the general features of the lensing mechanism have been recalled and explicit analytic relations have been given that are of help in understanding the main features. In particular Relations 5 and 6, describing the mapping in the linear approximation as a function of the lensing potential using a mapping matrix $\lambda$, are useful in practical cases to calculate image magnifications and to provide a qualitative picture of the lensing mechanism.

The study of the particular case of QSO RXJ0911 has illustrated how the problem becomes non linear when the source size becomes large enough for part of it to overlap the caustic. In such a case, the precise shape of the caustic in the source region becomes determinant in the construction of the images. The use of variables sensitive to the source size has been suggested, including the dependence of the relative image integrated fluxes. Typical quadruply imaged quasar configurations have been used as illustrations. The effect of strong lensing on a velocity gradient has been briefly considered.

\section{acknowledgements}
      Financial and/or material support from the Institute for Nuclear Science and Technology, National Foundation for Science and Technology Development (NAFOSTED), the World Laboratory, the French CNRS, the French Embassy in Hanoi, Rencontres du Vietnam and Odon Vallet fellowships is gratefully acknowledged.


\label{lastpage}

\end{document}